\documentclass{aa}
\usepackage{newtxtext,newtxmath}

\usepackage[T1]{fontenc}

\usepackage{natbib}
\usepackage{graphicx}
\usepackage{amsmath}
\usepackage{textcomp}
\usepackage{verbatim}
\usepackage{subcaption}
\usepackage{lscape}
\usepackage{placeins}
\usepackage{booktabs}
\usepackage{float}
\usepackage[version=4]{mhchem}
\usepackage{xcolor}
\newcommand{\changed}[1]{\textcolor{black}{#1}}

\title{Robustness of C/O and mean molecular weight under disequilibrium chemistry}

\author{
B. Burt\inst{1,2}\email{Bethany.Burt@lmu.de}
\and
K. Heng\inst{1,2,3}
\and
A. J. Cridland\inst{1,2}
\and
S.-M. Tsai\inst{4}
}

\institute{
Faculty of Physics, Ludwig Maximilian University, Scheinerstrasse 1, D-81679, Munich, Bavaria, Germany
\and
Munich Center for Geoastronomy, Ludwig Maximilian University, Theresienstrasse 41, D-80333, Munich, Bavaria, Germany
\and
University College London, Department of Physics \& Astronomy, Gower St, London, WC1E 6BT, United Kingdom
\and
Institute of Astronomy \& Astrophysics, Academia Sinica, Taipei 10617, Taiwan
}

\date{Received 14 July 2026}

\begin{document}

\abstract
{Transmission spectroscopy increasingly reduces exoplanet spectra to derived compositional metrics, such as C/O, O/H, and mean molecular weight ($\mu$), but these quantities do not carry molecular abundance uncertainties in the same way.}
{We \changed{assessed} the reliability of these metrics for hot Jupiters and sub-Neptunes.}
{We \changed{combined} analytic uncertainty propagation with photochemical \texttt{VULCAN} forward models and representative published retrieval constraints.}
{Propagating molecular abundance uncertainties produced substantially larger fractional uncertainties in elemental abundance ratios than in $\mu$. The fractional uncertainties in C/O and O/H were typically comparable to the inferred ratios, whereas the fractional uncertainty in $\mu$ was more than two orders of magnitude smaller in \ce{H2}/He-dominated atmospheres. Photochemistry and vertical mixing redistributed individual molecules but produced only minor changes in C/O and $\mu$ at the pressures probed by transmission spectroscopy.}
{C/O remains an important diagnostic of atmospheric chemistry and planetary formation history, but its interpretation requires careful uncertainty propagation. Mean molecular weight may provide a comparatively stable population-level diagnostic for linking observable atmospheres to bulk composition, especially across the sub-Neptune radius valley.}


\maketitle
\nolinenumbers
%%%%%%%%%%%%%%%%%%%%%%%%%%%%%%%%%%%%%%%%%%%%%%%%%%

%%%%%%%%%%%%%%%%% BODY OF PAPER %%%%%%%%%%%%%%%%%%

\section{Introduction}
Transmission spectroscopy is a primary route to atmospheric characterisation for transiting exoplanets, especially those with extended atmospheres and frequent transits. Wavelength-dependent transit depths arise because starlight passes through the planetary limb and acquires opacity signatures from gases and aerosols \citep{seagersasselov2000,brown2001,hubbard2001,benneke2012retrieval}. The James Webb Space Telescope (JWST) now provides broad wavelength coverage and high precision, enabling detections of multiple molecules in hot Jupiters, warm gas giants, and sub-Neptunes \citep{rustamkulov2023nirspecprism,bell2023wasp80,xue2024hd209458,madhusudhan2023k218b,benneke2024toi270d}. Improved data quality does not remove a central limitation: interpretation often compresses diverse spectra into a small set of compositional metrics whose reliability depends on modelling assumptions.

Retrieved abundances of common molecular opacity sources such as \ce{H2O}, \ce{CO}, \ce{CO2}, and \ce{CH4} often carry large fractional uncertainties. Across representative retrievals of hot Jupiters, warm Neptunes, and sub-Neptunes, reliably detected species typically have fractional abundance uncertainties of approximately $50$--$150\%$, while marginal detections and upper limits are substantially less constrained \citep{Lueber24,Bachmann25,Felix25}. Therefore, atmospheric retrievals commonly report derived quantities such as the carbon-to-oxygen ratio (C/O), elemental carbon abundance (C/H), and mean molecular weight $\mu$, all of which are calculated from uncertain molecular mixing ratios. Their uncertainties depend on wavelength coverage, spectral resolution, data quality, cloud treatment, and whether individual molecules are strongly required by the data or only moderately favoured in Bayesian model comparison \citep{chapman2017spectralcoverage,Felix25}. Typical thresholds for strong and moderate evidence are $\ln\mathcal{B}\gtrsim5$ and $\ln\mathcal{B}\sim2.5$, respectively \citep{madhusudhanseager2009,barstowheng2020,welbanksmadhusudhan2019,Lueber24}. Molecular abundance uncertainties do not propagate equally into C/O, C/H, and $\mu$. Elemental ratios depend on sums and ratios of trace species, whereas $\mu$ weights each species by its molecular mass and may be comparatively insensitive to uncertainty in a single trace molecule when the atmosphere is dominated by \ce{H2}. Therefore, the relevant question is not only what value is inferred, but also how strongly each derived metric amplifies or partially cancels the uncertainties in the retrieved molecular abundances.

Atmospheric processing may also affect these metrics differently. Assumptions about clouds, thermal structure, chemistry, reference pressure, and radius normalisation influence the inferred molecular abundances even when the observations are precise. Vertical mixing and photochemistry further modify the abundances at the low pressures probed by transmission spectroscopy. \changed{Therefore, a robust compositional metric should satisfy} two conditions: it should carry a small propagated uncertainty from the retrieved molecular abundances, and it should remain stable under plausible changes in atmospheric chemistry and transport.

C/O is widely used because it controls the distribution of carbon and oxygen among the dominant molecular carriers in hydrogen-dominated atmospheres and can strongly reshape the spectrum \citep{madhusudhan2012,Molliere2015,FortneyReview2018}. C/O is also interpreted as a tracer of planet formation through disk snowline chemistry \citep{oberg2011,Cridland2016a,Penzlin2024}. C/H provides a complementary absolute abundance scale. We define C/H as the elemental carbon abundance relative to hydrogen, thereby avoiding ambiguity with bulk metallicity.

Mean molecular weight, $\mu$, is motivated by the geometry of radiative transfer. In a hydrostatic atmosphere, the pressure scale height satisfies
$H \propto T\,\mu^{-1}\,g^{-1}$ where $T$ is temperature and $g$ is gravitational acceleration. \changed{Therefore, mean molecular weight helps} determine the amplitude of spectral features in transmission \citep{lecavelier2008,dewitseager2013}. This relation motivates the interpretation of muted spectra as evidence for atmospheres with high $\mu$ \citep{benneke2012retrieval}. Transmission spectra, however, lack an absolute normalisation and contain degeneracies among composition, reference pressure, and transit radius \citep{hengkitzmann2017,betremieuxswain2017,welbanksmadhusudhan2019}. Aerosols introduce another degeneracy because clouds and hazes can flatten spectral features and mimic the effect of high $\mu$ \citep{fortney2005,bean2010,millerricci2010,bennekeseager2013}. Precise observations are required to distinguish clouds at high altitude from atmospheres with high $\mu$ \citep{kreidberg2014}.

Sub-Neptunes strengthen the motivation for considering $\mu$ in population studies. \citet{hengowentian2025} predicted a gradient in $\mu$ across the observed radius valley \citep{fulton2017,owenwu2017}, driven primarily by interior oxygen fugacity rather than atmospheric carbon enrichment. This prediction suggests that $\mu$ may encode interior conditions. It also exposes a modelling challenge: connecting bulk or outgassed composition to an observable atmosphere requires tracking how chemistry, mixing, and irradiation process molecular species at the pressures probed by transmission spectroscopy.

Photochemistry provides a direct test of metric robustness because it alters atmospheric composition at low pressures. Transport and photochemistry modify vertical abundance profiles and shift the observable composition away from local thermochemical equilibrium \citep{prinnbarshay1977,coopershowman2006,moses2011,venot2012,huseager2014}. The detection of the photochemical product \ce{SO2} in WASP-39b demonstrated that disequilibrium chemistry can directly affect observed spectra \citep{rustamkulov2023nirspecprism,Tsai23}. \changed{Therefore, C/O, C/H, and $\mu$ may respond} differently to the same physical perturbation.

Here we assessed the reliability of $\mu$ and C/O as compositional metrics in transmission spectroscopy for sub-Neptunes and hot Jupiters. First, we analytically propagated uncertainties in retrieved volume mixing ratios through a chemical equilibrium network containing carbon, hydrogen, and oxygen to compare the intrinsic uncertainties in $\mu$, C/O, and C/H. Second, we used {\tt VULCAN} forward models to quantify how these metrics responded to photochemistry, vertical mixing, metallicity, irradiation, and planetary type \citep{tsai2017vulcan,tsai2021vulcancomparative}. Finally, we applied the same uncertainty propagation framework to molecular abundances from published retrievals. Our goal was not to replace elemental ratios with $\mu$, but to determine when $\mu$ was robust and when it depended strongly on the atmospheric model. By benchmarking $\mu$ against C/O and C/H across hot Jupiter and sub-Neptune regimes, we provided a framework for interpreting JWST transmission spectra in population studies.

\section{Methods}
\label{sec:methods}

\subsection{Intrinsic uncertainty propagation in the C--H--O system}

We \changed{used} a hydrogen-dominated C--H--O atmosphere to estimate the intrinsic uncertainty associated with each compositional metric. The calculation \changed{used} an equilibrium chemistry network containing $\mathrm{CH_4}$, $\mathrm{H_2O}$, $\mathrm{CO}$, $\mathrm{CO_2}$, and $\mathrm{C_2H_2}$. These species \changed{captured} the dominant carbon- and oxygen-bearing molecules in hot hydrogen-rich atmospheres and \changed{matched} the C--H--O system used in previous analytic equilibrium-chemistry studies \citep{Heng16,Heng16b}.

We \changed{specified} the temperature $T$, pressure $P$, elemental ratio C/O, and elemental abundances $n_{\mathrm{O}}$ and
$n_{\mathrm C}$. We \changed{fixed} $n_{\mathrm{O}}=5\times10^{-4}$ and \changed{sampled} $T$, $P$, and C/O over 100 realisations using $T\sim\mathcal{U}(800,2500)\,\mathrm{K}$,
$\log(P/\mathrm{bar})\sim\mathcal{U}(-3,2)$, and
$\mathrm{C/O}\sim\mathcal{U}(0.1,1.0)$. We then calculated the carbon abundance $n_{\mathrm C}$ from the selected C/O ratio according to
\begin{equation}
n_{\mathrm C} = \text{C/O} \times n_{\mathrm O}.
\end{equation}
The temperature and C/O ranges \changed{span} the hot-Jupiter and sub-Neptune regimes considered in the forward models. The pressure range \changed{was} broader than the nominal transmission region used later because this calculation \changed{sampled} the equilibrium compositions available across a vertically mixed atmosphere, rather than the composition at a single pressure level. In this sense, the sampled pressures \changed{approximated} the thermochemical reservoir that may be transported into the observable atmosphere by vertical mixing.

For each sampled state, we \changed{computed} equilibrium constants and molecular mixing ratios $X_i$ by solving the polynomial system for the C--H--O network. The temperature-dependent Gibbs free-energy differences,
$\Delta G_i(T)$, were taken from the National Institute of Standards
and Technology--Joint Army-Navy-Air Force (NIST--JANAF)
Thermochemical Tables and linearly interpolated in temperature, following \cite{Heng16} and \cite{Heng16b}. The NIST-JANAF tables provide recommended temperature-dependent thermochemical quantities, including Gibbs free energies and equilibrium constants, for gas-phase species (National Institute of Standards and Technology 1998).

For each realisation, equilibrium mixing ratios \changed{were computed} for
\ce{H2O}, CO, \ce{CO2}, and \ce{CH4}, with \ce{H2} \changed{obtained} from number
conservation. Elemental budgets \changed{were defined} as
$N_{\rm O}=X_{\rm CO}+2X_{\rm CO_2}+X_{\rm H_2O}$ and
$N_{\rm C}=X_{\rm CO}+X_{\rm CO_2}+X_{\rm CH_4}$, so that
O/H$=N_{\rm O}/(2X_{\rm H_2})$, C/O$=N_{\rm C}/N_{\rm O}$, and
$\mu=2X_{\rm H_2}+18X_{\rm H_2O}+28X_{\rm CO}+44X_{\rm CO_2}+16X_{\rm CH_4}$.

To isolate the intrinsic error amplification of each compositional metric, we \changed{assigned} every molecular abundance the same fractional uncertainty, $\delta$: 
\begin{equation}\label{eqn:frac_sigma}
    \sigma_{X_i}=\delta X_i .
\end{equation}
We \changed{treated} the abundance errors as independent and uncorrelated between species. This assumption \changed{isolated} how the algebraic form of each metric propagates molecular abundance errors, and \changed{was not intended} to reproduce the covariance structure of a full atmospheric retrieval. Under this assumption, the fractional uncertainty in a weighted sum
$A=\sum_i a_iX_i$ is
\begin{equation}
    \frac{\sigma_A}{A}
    =
    \delta\,
    \frac{\sqrt{\sum_i a_i^2X_i^2}}
         {\sum_i a_iX_i},
\end{equation}
and for a ratio $z=x/y$ of independent quantities,
\begin{equation}
    \frac{\sigma_z}{z}
    =
    \sqrt{\left(\frac{\sigma_x}{x}\right)^2+\left(\frac{\sigma_y}{y}\right)^2}.
\end{equation}
We \changed{neglected} uncertainty in $X_{\rm H_2}$, which is not directly measured.
\begin{table*}[h]
\centering
\caption{Stellar and planetary parameters, along with retrieved molecular abundances, for the four case studies.}
\label{tab:combined}
\begin{tabular}{lcccc}
\hline
\hline
Parameter / species
    & WASP-39b
    & HD~209458~b
    & TOI-270~d
    & HAT-P-26~b \\
\hline
Model
    & PRISM grey
    & Clouds + informed CO prior
    & Fiducial
    & Best-fit \\
Retrieval reference
    & \citet{Lueber24}
    & \citet{Bachmann25}
    & \citet{Felix25}
    & \citet{Gressier25} \\
$T_{\rm eff,\star}$ (K)
    & $5485 \pm 50$
    & 6030
    & $3506 \pm 70$
    & $5104^{+89}_{-103}$ \\
Mass of planet ($M_{\rm Jup}$)
    & $0.281 \pm 0.032$
    & $0.682^{+0.014}_{-0.015}$
    & $0.0150 \pm 0.0014$
    & $0.0585 \pm 0.0069$ \\
Radius of planet ($R_{\rm Jup}$)
    & $1.279 \pm 0.040$
    & $1.359^{+0.016}_{-0.019}$
    & $0.1903 \pm 0.0052$
    & $0.550 \pm 0.010$ \\
Stellar reference
    & \citet{Mancini18}
    & \citet{Rosenthal21}
    & \citet{VanEylen21}
    & \citet{Gressier25} \\
Planetary reference
    & \citet{Mancini18}
    & \citet{Bonomo17}
    & \citet{VanEylen21}
    & \citet{Wakeford17} \\
\hline
$\log{\ce{H2O}}$
    & $-3.10^{+0.20}_{-0.19}$
    & $-2.80^{+0.42}_{-0.41}$
    & $-4.1^{+1.7}_{-5.0}$
    & $-2.07^{+0.39}_{-0.59}$ \\
$\log{\ce{CO}}$
    & $-2.85^{+0.17}_{-0.28}$
    & $-4.15^{+0.51}_{-0.30}$
    & $< -4$
    & $-5.3^{+2.4}_{-4.1}$ \\
$\log{\ce{CO2}}$
    & $-4.34^{+0.24}_{-0.22}$
    & $-6.16^{+0.38}_{-0.39}$
    & $-1.29^{+0.17}_{-0.19}$
    & $-2.91^{+0.56}_{-0.71}$ \\
$\log{\ce{CH4}}$
    & $\ldots$
    & $-9.4^{+2.2}_{-2.3}$
    & $-1.72^{+0.28}_{-0.29}$
    & $< -5.49$ \\
\hline
O/H
    & $0.00232^{+0.00078}_{-0.00071}$
    & $0.0017^{+0.0026}_{-0.0010}$
    & $0.103^{+0.049}_{-0.036}$
    & $0.0111^{+0.014}_{-0.0066}$ \\
C/O
    & $0.63^{+0.36}_{-0.37}$
    & $0.043^{+0.099}_{-0.071}$
    & $0.76^{+0.41}_{-0.42}$
    & $0.11^{+0.32}_{-0.17}$ \\
$\mu$ (amu)
    & $2.572^{+0.020}_{-0.020}$
    & $2.538^{+0.051}_{-0.018}$
    & $5.64^{+1.4}_{-0.91}$
    & $2.72^{+0.40}_{-0.16}$ \\
\hline
\end{tabular}
\tablefoot{The ``Retrieval reference'' row identifies the source of the atmospheric model and molecular abundances for each planet. The ``Stellar reference'' and ``Planetary reference'' rows identify the sources of $T_{\rm eff,\star}$ and of $M_{\rm p}$ and $R_{\rm p}$, respectively. For HAT-P-26 b, the planetary mass and its uncertainty are taken from \citet{Wakeford17}, while the reference radius is retrieved by \citet{Gressier25}. The O/H, C/O, and $\mu$ values were calculated in this work from the reported molecular abundances above the final horizontal division. Values without uncertainties were adopted as fixed inputs without quoted uncertainties in the corresponding retrieval study.}
\end{table*}
Applying these expressions gives the propagated fractional uncertainties
\begin{equation} \label{eqn:oh_error}
    \frac{\sigma_{\text{O/H}}}{\text{O/H}}
    \approx
    \delta\,
    \frac{\sqrt{X_{\rm CO}^2+4X_{\rm CO_2}^2+X_{\rm H_2O}^2}}
         {X_{\rm CO}+2X_{\rm CO_2}+X_{\rm H_2O}},
\end{equation}
\begin{equation}\label{eqn:co_error}
    \frac{\sigma_{\text{C/O}}}{\text{C/O}}
    =
    \delta\sqrt{
    \frac{X_{\rm CO}^2+X_{\rm CO_2}^2+X_{\rm CH_4}^2}
         {(X_{\rm CO}+X_{\rm CO_2}+X_{\rm CH_4})^2}
    +
    \frac{X_{\rm CO}^2+4X_{\rm CO_2}^2+X_{\rm H_2O}^2}
         {(X_{\rm CO}+2X_{\rm CO_2}+X_{\rm H_2O})^2}
    },
\end{equation}
and
\begin{equation}\label{eqn:mmw_error}
    \frac{\sigma_\mu}{\mu}
    =
    \delta\,
    \frac{
    \sqrt{324X_{\rm H_2O}^2+784X_{\rm CO}^2+1936X_{\rm CO_2}^2+256X_{\rm CH_4}^2}
    }{
    2X_{\rm H_2}+18X_{\rm H_2O}+28X_{\rm CO}+44X_{\rm CO_2}+16X_{\rm CH_4}
    }.
\end{equation}
The first expression uses only the uncertainty in the O/H numerator
because the atmosphere is \ce{H2}-dominated. The second follows from
C/O$=N_{\rm C}/N_{\rm O}$, and the coefficients in the third equation are
the squared molecular weights, e.g.\ $18^2=324$.

\noindent To assess the robustness of the thermochemical uncertainty framework to disequilibrium effects, we \changed{applied} an idealised photochemical perturbation to the equilibrium abundances before propagating errors. This perturbation \changed{was not intended} to reproduce a complete photochemical network. It \changed{provided} a controlled test of how photochemistry-induced shifts in the abundances of the major observable C--H--O species \changed{modified} the inferred fractional uncertainties. We \changed{approximated} the net redistribution among observable species using two simplified stoichiometric proxies representative of the dominant redistribution among the observable C--H--O molecular reservoirs in warm \ce{H2}-dominated atmospheres. \citep[e.g.][]{Moses11,Tsai17}.

The first proxy \changed{represented} the net oxidation of CO to CO$_2$ driven by water photolysis. Bolded species are those which appear in the net balanced equation.
\begin{alignat}{2}
\ce{\bf{H_2O} &-> H\cdot \text{+} OH\cdot}
     & \text{photolysis} \\
\ce{OH\cdot \text{+} \textbf{CO} &-> H\cdot \text{+} \bf{CO_2}}
    & & \text{CO oxidation} \\
\ce{H\cdot \text{+} H\cdot &-> \bf{H_2}}
    & & \text{radical recombination}
\end{alignat}
which is represented by the net stoichiometric conversion
\begin{align}\label{eqn:CO_oxidation}
R_1:\quad
\ce{CO + H2O -> CO2 + H2}.
\end{align}

The second proxy \changed{represented} oxidation of \ce{CH4} to CO through intermediate hydrocarbons and oxygen-bearing species,
\begin{alignat}{2}\label{eqn:CH4_oxidation}
\ce{\bf{CH_4} + H\cdot &-> CH3\cdot $+$ \bf{H_2}}
     & \text{H abstraction} \\
\ce{CH3\cdot $+$ OH\cdot &-> CH2OH\cdot $+$ H\cdot}
     & \text{C--O bond formation} \\
\ce{CH2OH\cdot $+$ M &-> H2CO + H\cdot $+$ M}
  & \text{H$_2$CO formation} \\
\ce{H2CO + H\cdot &-> HCO\cdot $+$ \bf{H_2}}
    & \text{H$_2$CO destruction} \\
\ce{HCO\cdot $+$ M &-> H\cdot $+$ \bf{CO} + M}
     & \text{CO formation} \\
\ce{H\cdot $+$ H2O &-> OH\cdot $+$ \bf{H_2}}
    & \text{radical recycling}
\end{alignat}
The reactions \changed{gave} the net conversion
\begin{equation}
R_2:\quad\ce{CH4 + H2O -> CO + 3H2}
\end{equation}
In photochemical regions, \ce{H2O} photolysis supplies H and OH radicals that initiate or facilitate methane destruction. The net conversion treats \ce{H2O} as the oxygen source incorporated into CO rather than as a unique reactive intermediate. For each Monte Carlo realisation, we first \changed{computed} the equilibrium abundances of \ce{CH4}, \ce{H2O}, CO, \ce{CO2}, and \ce{C2H2}. We then \changed{converted} a fraction \(f=\{0.10,0.50,1.00\}\) of the relevant reactant pool, subject to stoichiometric limits that prevent negative abundances. R1 \changed{was limited} by the available CO and \ce{H2O}, while R2 \changed{was limited} by the available \ce{CH4} and \ce{H2O}. We \changed{evaluated} three perturbation cases: R1 alone, R2 alone, and R2 followed by R1, with the same value of \(f\) applied to both reactions in the combined case. The R2 conversion \changed{increased} the total number of molecules because two reactant molecules produce four product molecules. \changed{Therefore, we renormalised} all molecular mixing ratios after applying R2. \(\ce{C2H2}\) \changed{did not participate} in either proxy reaction, and the \(\ce{H2}\) mixing ratio \changed{was recovered} from mixing-ratio closure. The unperturbed equilibrium solution (\(f=0\)) \changed{provided} the thermochemical reference case. We then \changed{re-evaluated} Eqs.~(\ref{eqn:oh_error})--(\ref{eqn:mmw_error}) using the perturbed abundances. 

This procedure \changed{did not introduce} an additional photochemical error term, but \changed{quantified} how plausible photochemistry-induced redistribution among the observable molecular reservoirs \changed{modified} the propagation of thermochemical uncertainties.

\subsubsection{Application to published retrievals}

To evaluate the analytic uncertainty framework using representative observational data, we selected four published atmospheric retrievals spanning H$_2$/He-dominated hot Jupiters and a metal-rich sub-Neptune (Table~\ref{tab:combined}). Retrieved molecular abundances of \ce{H2O}, CO, \ce{CO2}, and \ce{CH4} were compiled directly from the published posterior constraints. These abundances were used to calculate the corresponding O/H, C/O, and mean molecular weight using the framework described above. Uncertainties on the derived quantities were obtained by propagating the reported molecular abundance uncertainties through Eqns.~\eqref{eqn:oh_error}--\eqref{eqn:mmw_error}. This comparison \changed{provided} an observational test of the analytic predictions across a range of atmospheric compositions and retrieval precisions.

\subsection{Forward-model population synthesis}\label{sec:method_sampling}

We \changed{generated} population-level predictions for C/O and $\mu$ by constructing a suite of atmospheric models for four sub-populations of sub-Neptune- and hot-Jupiter-like exoplanets. The sub-populations \changed{varied} in planetary class, host-star spectral type, and chemical network (Table~\ref{tab:subpop_params}).  

We \changed{computed} these models with the one-dimensional photochemical kinetics code \texttt{VULCAN} \citep{Tsai17, Tsai21}, which solves the coupled continuity equations for chemical species under the combined effects of thermochemistry, photolysis, and vertical mixing.

For each model, the \changed{pressure--temperature profile was calculated} using the \citet{Guillot10} formalism, following the analytical solution of \citet{Heng16b}, which allows for scattering. The irradiation temperature \changed{was calculated} from the stellar temperature and semi-major axis, with the stellar radius fixed to $1\,R_{\odot}$ for all model groups.
Although this value is larger than a representative M-dwarf radius,
we adjusted $a$ to reproduce each target $T_{\rm eq}$. Consequently,
$a$ scaled linearly with $R_\star$, leaving the ratio $R_\star/a$,
and hence the stellar-flux dilution factor $(R_\star/a)^2$, unchanged.
\changed{Therefore, the assumed stellar radius did not affect} the irradiation
received by the model atmosphere at fixed $T_{\rm eq}$; it only
produced nominal semi-major axes larger than those expected for a
typical M-dwarf radius. The longwave and shortwave opacities \changed{were fixed} to 0.01 and $0.005\,\text{cm}^2\,\text{g}^{-1}$, respectively.

The \changed{pressure--temperature profile was coupled} to the relevant chemical network, as listed in Table \ref{tab:subpop_params}, and the model \changed{was evolved} to a converged steady-state abundance profile. To control the timestep size, we \changed{used} an adaptive relative tolerance, setting the initial relative tolerance to $1$, with lower and upper limits of $0.02$ and $2.5$, respectively.
% He/H, P_t, P_b, no. layers? 

\subsubsection{Sampled parameters}
To explore the effect of chemical composition, we \changed{varied} the input metallicity, parameterised by O/H, and the elemental C/O ratio. The input values \changed{were generated} using uniform \changed{Monte Carlo} sampling. We sampled C/O uniformly between 0.1 and 1.0 and sampled O/H from a
log-uniform distribution between $5\times10^{-5}$ and
$5\times10^{-2}$. We then calculated C/H from the sampled C/O and O/H
values.
Finally, to incorporate the effect of the \changed{pressure--temperature structure}, we \changed{varied} the equilibrium temperature, $T_{\mathrm{eq}}$, from $800\,$K to $2500\,$K. Since \texttt{VULCAN} does not use $T_{\mathrm{eq}}$ as a direct input parameter, this variation \changed{was imposed} by adjusting the semi-major axis, $a$. We calculated $a$ assuming complete heat
redistribution and a Bond albedo of $A_{\rm B}=0$. The zero-albedo
assumption corresponds to complete absorption of the incident stellar
radiation and provided a consistent irradiation convention across the
generic model grid without introducing an additional poorly
constrained parameter.

Using the three variable parameters, O/H, C/O, and $T_{\mathrm{eq}}$, we randomly \changed{generated} 125 unique parameter combinations. For each combination, we \changed{constructed} four classes of model: thermochemical+low-mixing, thermochemical+high-mixing, photochemical+low-mixing, and photochemical+high-mixing. The photochemical models \changed{included} photolysis in the \texttt{VULCAN} chemical solver, whereas the thermochemical models \changed{excluded} photolysis and \changed{evolved} the atmosphere under thermal chemistry and vertical transport.
\begin{table*}[t]
\centering
\caption{{\tt VULCAN} input parameters adopted for each atmospheric model group.}
\label{tab:subpop_params}

\begin{tabular}{cccccccc}
\hline
\hline
&
\multicolumn{2}{c}{Stellar parameters} &
\multicolumn{5}{c}{Planetary parameters} \\
\cmidrule(lr){2-3}
\cmidrule(lr){4-8}
Model group &
$T_\star$ (K) &
Spectral type &
Planet type &
Chemical network &
Mass ($M_{\rm Jup}$) &
Radius ($R_{\rm Jup}$) &
$g_{\rm surface}$ ($\mathrm{cm\,s^{-2}}$) \\
\hline

A & 5772 & G & Sub-Neptune & CHO
  & 0.0186 & 0.196 & 1200 \\

B & 5772 & G & Hot Jupiter & CHO
  & 1.03 & 1.13 & 2000 \\

C & 5772 & G & Sub-Neptune & CHONS
  & 0.0186 & 0.196 & 1200 \\

D & 3473 & M & Sub-Neptune & CHO
  & 0.0186 & 0.196 & 1200 \\

\hline
\end{tabular}
\end{table*}

Vertical mixing \changed{was parameterised} by a constant eddy diffusion coefficient, $K_{\mathrm{zz}}$. Since $K_{\mathrm{zz}}$ remains poorly constrained observationally, it is commonly treated as a free parameter or estimated from simplified transport prescriptions in one-dimensional photochemical models \citep[e.g.][]{Allen81,Komacek19}. We \changed{adopted} two constant values, $K_{\mathrm{zz}} = 10^{6}$ and $10^{10}\,\mathrm{cm^{2}\,s^{-1}}$, which \changed{spanned} four orders of magnitude and \changed{fell} within the range of transport efficiencies explored in previous disequilibrium chemistry studies and inferred from global climate model (GCM)-based estimates of vertical mixing \citep[e.g.][]{Baxter21,Charnay15}. The lower value \changed{was} representative of weak-mixing cases used in sub-Neptune photochemical models, while the upper value \changed{was} comparable to the strong-mixing cases and upper-atmosphere eddy diffusion coefficients adopted in one-dimensional hot-Jupiter and {\tt VULCAN} photochemical models \citep[e.g.][]{MillerRicciKempton12,Tsai23}. \changed{Therefore, these two cases provided} a simple framework for assessing the sensitivity of transmission-region chemistry to vertical transport while minimizing the dimensionality of the parameter space. The four model classes applied to the 125 sampled parameter combinations \changed{yielded} 500 models per sub-population, and 2000 models in total.

\subsubsection{Metric calculation}

From each converged model, we \changed{calculated} $\mu$ and C/O over the pressure range $10^{-3}$--$10^{-1}\,$bar, corresponding to the atmospheric region typically probed by transmission spectroscopy.

The mean molecular weight \changed{was computed} as

\begin{equation}\label{eqn:mu}
    \mu = \sum_i X_i \mu_i,
\end{equation} 

\noindent while elemental ratios \changed{were computed} as

\begin{equation}\label{eqn:co}
    \text{C/O} = \frac{\sum_{i \in \mathrm{C}} X_i \, N_{i,\mathrm{C}}}{\sum_{i \in \mathrm{O}} X_i \, N_{i,\mathrm{O}}},
\end{equation}\\
 where $N_{i,\mathrm{C}}$ and $N_{i,\mathrm{O}}$ are the numbers of carbon and oxygen atoms in species $i$, respectively. These quantities \changed{were evaluated} at each pressure level and \changed{averaged} over the selected pressure range.

We \changed{quantified} robustness by comparing matched model pairs that differed only in chemistry or in $K_{zz}$. For chemistry, we \changed{subtracted} the $\mu$ of the thermochemical model from the photochemical $\mu$ value and \changed{normalised} by the thermochemical value. We \changed{made} this comparison separately for the low- and high-mixing pairs. For vertical mixing, we \changed{subtracted} the low-mixing value from the high-mixing value and \changed{normalised} by the low-mixing value, holding the chemistry fixed. %Figure~\ref{fig:main_panels} shows these four matched comparisons.

\begin{equation}\label{eqn:delta_chemical}
    \Delta \Theta_{\mathrm{chem}} = \frac{\Theta_{\mathrm{photochemical}}- \Theta_{\mathrm{thermochemical}}}{\Theta_{\mathrm{thermochemical}}}
\end{equation}

\begin{equation}\label{eqn:delta_mixing}
    \Delta \Theta_{\mathrm{mix}} = \frac{\Theta_{\mathrm{high~mixing}}- \Theta_{\mathrm{low~mixing}}}{\Theta_{\mathrm{low~mixing}}}
\end{equation}\\
where $\Theta$ is the pressure-averaged transmission-region value of either C/O or $\mu$.

\section{Results}\label{sec:results}

\subsection{Intrinsic uncertainty is lower for $\mu$ than for elemental ratios}\label{sec:results-error_analysis}

\begin{figure}
    \centering
    \includegraphics[width=\linewidth]{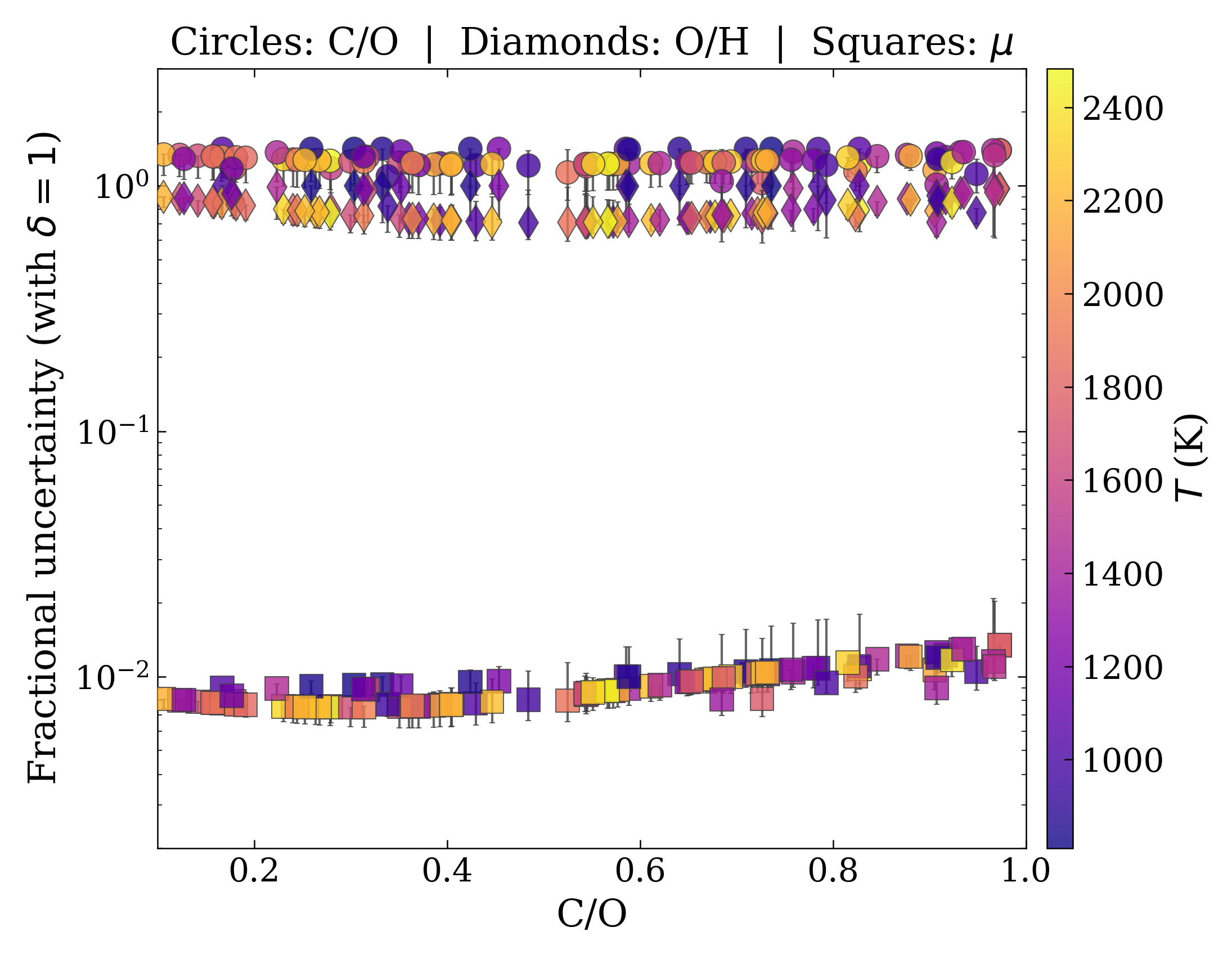}
     \caption{Propagated fractional uncertainties on C/O (circles), O/H (diamonds), and mean molecular weight $\mu$ (squares) for equilibrium C--H--O compositions sampled in $T$--$P$--C/O space, assuming a common per-species abundance uncertainty $\delta=1$. Marker colour indicates temperature. Vertical error bars indicate how far photochemical perturbations can raise or lower the propagated fractional uncertainty relative to the thermochemical baseline (maximum shift in each direction across R1 and R2 applied simultaneously at reactant conversion fraction of 0.1, 0.5, and 1.0).}
    \label{fig:error_result}
\end{figure}

Analytic uncertainty propagation predicts substantially larger fractional uncertainties for elemental abundance ratios than for mean molecular weight (Fig.~\ref{fig:error_result}). The propagated uncertainties in O/H, C/O, and $\mu$ were evaluated using Eqns.~\eqref{eqn:oh_error}--\eqref{eqn:mmw_error} over the sampled $T$--$P$--C/O grid. Assuming a common per-species abundance uncertainty of $\delta=1$, the median fractional uncertainty on C/O is approximately 1.26, while O/H exhibits a similarly large uncertainty of $\sigma_{\mathrm{O/H}}/\mathrm{(O/H)}\approx0.7$--$1.0$. In contrast, the propagated uncertainty on $\mu$ remains more than two orders of magnitude smaller, with a median value of $9\times10^{-3}$, and varies only weakly across the sampled parameter space.

This difference in error propagation arises from the different roles of the trace molecular species in each derived quantity. Mean molecular weight is dominated by the \ce{H2}/He background, so uncertainties in the trace C--H--O species perturb only a small fraction of the total atmospheric mass. Elemental abundance ratios are constructed entirely from those same trace species, allowing molecular abundance uncertainties to propagate directly into C/O and O/H. The analytic framework also predicts different sensitivities to atmospheric metallicity. As heavy molecules comprise an increasingly large fraction of the bulk atmosphere, the uncertainty on $\mu$ increases as the dominance of the H$_2$/He background diminishes. In contrast, the fractional uncertainty on C/O is comparatively insensitive to metallicity because both the numerator and denominator scale with the abundances of the same heavy-element carriers, leaving the propagated uncertainty primarily controlled by the relative precision of the retrieved molecular abundances.

The photochemical perturbations preserve the contrast predicted by the analytic calculation
(Fig.~\ref{fig:error_result}). The vertical
bars show the maximum decreases and increases relative to the baseline
thermochemical calculation when reactions R1 (Eqn.~\ref{eqn:CO_oxidation}) and R2 (Eqn~\ref{eqn:CH4_oxidation}) were applied
simultaneously, using the same conversion fraction,
$f=0.10$, $0.50$, or $1.00$, for both reactions. In all cases, the propagated
uncertainties in C/O and O/H remain of order unity, whereas those on $\mu$
remain more than two orders of magnitude smaller. The photochemical bars remain within the order of magnitude of the thermochemical calculations, showing that substantial
photochemistry-induced redistribution among the observable C--H--O reservoirs
does not overturn the contrast between elemental ratios and mean molecular
weight.

\subsection{Derived atmospheric metrics are robust to disequilibrium chemistry}

\begin{figure}
    \centering
    \includegraphics[width=\linewidth]{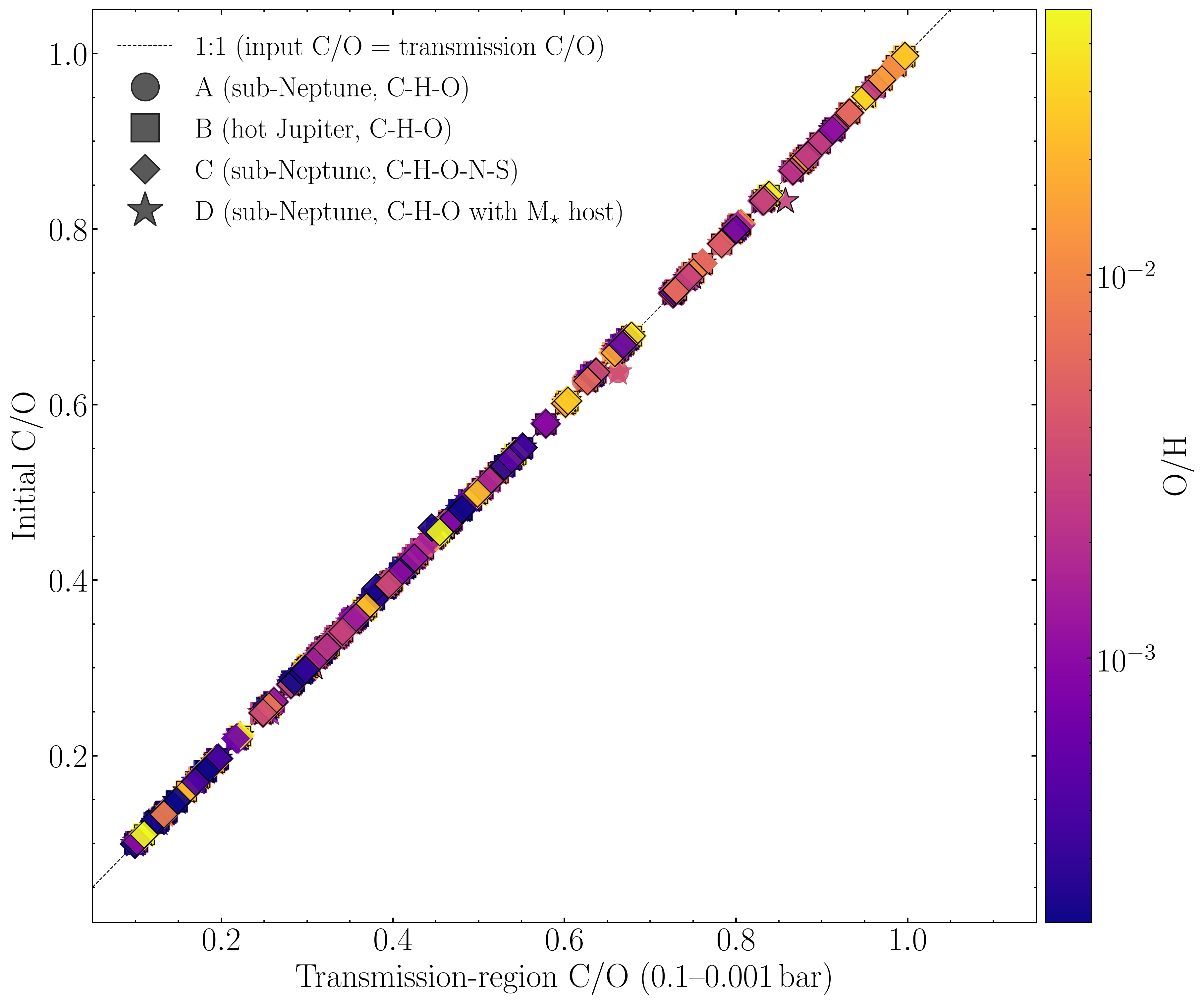}
    \caption{Input C/O compared with the transmission-region C/O recovered from converged photochemical models. The dashed red line marks one-to-one agreement. Models near the line preserve the input elemental ratio in the observable pressure range.}
    \label{fig:co_input_output}
\end{figure}

Photochemistry and vertical mixing do not introduce a systematic bias into the transmission-region C/O (Fig.~\ref{fig:co_input_output}). Across all model populations, the transmission-region C/O closely follows the input elemental ratio under thermochemical equilibrium, photochemical equilibrium, and both weak- and strong-mixing conditions. All models lie close to the one-to-one relation, indicating that disequilibrium chemistry does not systematically shift the inferred C/O away from its equilibrium value.

\begin{figure*}
    \centering
    \includegraphics[width=\linewidth]{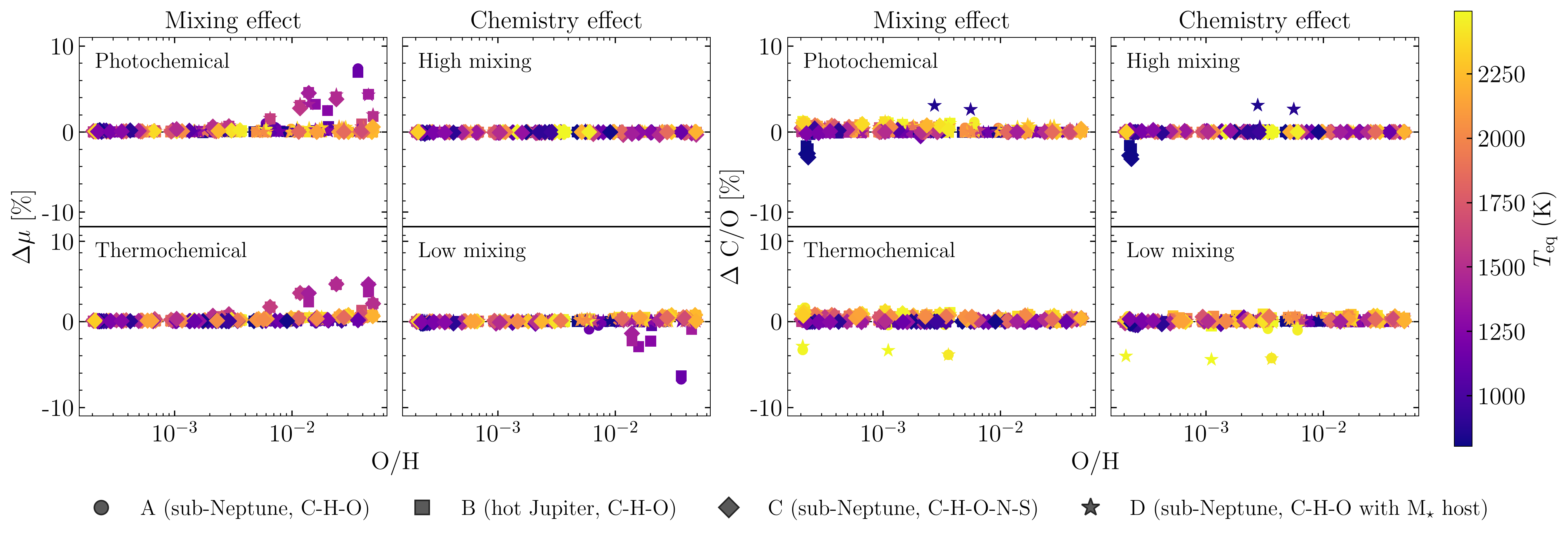}
    \caption{Fractional change in transmission-region mean molecular weight (left) and C/O (right) across matched model pairs. The left column in each subplot compares the effect of mixing in thermochemical and photochemical models (Eqn.~\ref{eqn:delta_mixing}), and the right column compares the effect of photochemistry in low- and high-mixing models (Eqn.~\ref{eqn:delta_chemical}). Values are plotted against bulk O/H, colour denotes $T_{\mathrm{eq}}$, and marker shape denotes model group. Most models remain within a few percent of zero, indicating that $\mu$ and C/O are weakly sensitive to photochemistry and vertical mixing across the sampled grid.}
    \label{fig:main_panels}
\end{figure*}

The residual changes in transmission-region C/O are also small (Fig.~\ref{fig:main_panels}). Relative to the thermochemical solution, most models differ by less than $1\%$, irrespective of photochemistry or vertical mixing ($\gtrsim98\%$ of chemistry pairs and $\gtrsim95\%$ of mixing pairs). The few outliers occur at the extremes of the temperature and abundance ranges, where the averaged C/O becomes sensitive to small changes in a limited number of trace molecular reservoirs.

The high-temperature outliers occur in thermochemical mixing pairs with $T_{\rm eq}\gtrsim2000~{\rm K}$ and low-to-intermediate O/H, approximately ${\rm O/H}\sim2\times10^{-4}$--$4\times10^{-3}$. These models reach $\Delta{\rm C/O}\simeq-3$ to $-4\%$, with high mixing producing lower C/O at pressures probed by transmission spectroscopy than low mixing. In these thermochemical atmospheres where C and O are carried mainly by \ce{H2O} and CO, increasing $K_{\rm zz}$ modestly increases the \ce{H2O} abundance and lowers the averaged C/O over the same pressure range. The absolute abundance changes are small, but the elemental ratio is sensitive because most of the carbon and oxygen are carried by only a few species.

A second group appeared in cool, extremely oxygen-poor photochemical
model pairs, with $T_{\rm eq}\sim800\,\mathrm{K}$ and
$\mathrm{O/H}\sim2\times10^{-4}$. These models showed the same response
as the hot thermochemical outliers: high mixing lowered C/O by
approximately $2$--$3\%$. However, the controlling chemical reservoirs
differed. Under photochemistry and high $K_{\rm zz}$, carbon was
redistributed from hydrogen cyanide (\ce{HCN}), partly into carbon
disulfide (\ce{CS2}) and methane (\ce{CH4}). The residual CO abundance
decreased substantially, whereas the \ce{H2O} abundance increased only
slightly. Ammonia (\ce{NH3}) increased concurrently as nitrogen
followed the same depletion of \ce{HCN}. Therefore, in this
oxygen-poor regime, C/O was sensitive to the redistribution of carbon
and oxygen among a small number of chemical reservoirs.

The remaining C/O outliers were confined to cool M-star models at
intermediate O/H, with $T_{\rm eq}\simeq840$--$880\,\mathrm{K}$ and
$\mathrm{O/H}\simeq3\times10^{-3}$--$6\times10^{-3}$. Unlike the
previous groups, high mixing increased the transmission-region C/O by
approximately $2$--$3\%$. In these models, carbon was redistributed
from methane into benzene (\ce{C6H6}), whereas oxygen remained
primarily in \ce{H2O}. This redistribution increased the carbon
inventory relative to the oxygen retained in \ce{H2O}, producing a
mixing response opposite to those of the hot and oxygen-poor outliers.

Mean molecular weight is similarly robust to disequilibrium chemistry for most models (Fig.~\ref{fig:main_panels}). Across the grid, $\mu$ is essentially unchanged for $\gtrsim93\%$ of mixing pairs. The exceptions are concentrated at high oxygen abundance and cool-to-intermediate equilibrium temperature, with ${\rm O/H}\simeq10^{-2}$--$5\times10^{-2}$ and $T_{\rm eq}\simeq1100$--$1550~{\rm K}$. In this regime, high mixing raises the transmission-region $\mu$ by approximately $2$--$7\%$ relative to low mixing.

These outliers occur because the transmission region becomes sensitive to vertical redistribution among the dominant heavy molecular carriers. In low-$K_{\rm zz}$ photochemical models, the $10^{-1}$--$10^{-3}{\rm bar}$ region remains vertically stratified, with \ce{H2O} and \ce{CH4} enhanced at higher pressures and CO more abundant at lower pressures. Strong eddy mixing homogenises this structure toward an \ce{H2O}- and \ce{CH4}-rich, CO-poor composition across the averaged pressure range, increasing the pressure-averaged $\mu$. Similar outliers in the thermochemical mixing comparisons arise from mixing-dependent quenching among the same molecular reservoirs, which also shifts the transmission region toward higher \ce{H2O} and \ce{CH4} and lower CO.

The individual molecular abundances exhibit substantially larger disequilibrium effects than the derived atmospheric metrics. Weakly mixed atmospheres develop larger photochemical departures than strongly mixed atmospheres, yet these molecular abundance changes produce only negligible changes in C/O and $\mu$.

\subsection{Published atmospheric retrievals support the analytic error-propagation pattern}

The benchmark retrievals reproduce the uncertainty hierarchy predicted by the analytic framework and preserved by the disequilibrium models (Table~\ref{tab:combined}). Despite spanning H$_2$/He-dominated hot Jupiters and a metal-rich sub-Neptune, all four retrievals exhibit substantially larger fractional uncertainties in C/O than in $\mu$.

The three H$_2$/He-dominated atmospheres (WASP-39b, HD~209458~b, and HAT-P-26~b) exhibit fractional uncertainties on C/O ranging from approximately $60\%$ to more than $200\%$, despite molecular abundance uncertainties of tens of percent to order unity. In contrast, the corresponding uncertainties on $\mu$ remain below $\sim10\%$ and are close to $1\%$ for WASP-39b and HD~209458~b.

TOI-270~d departs from this behaviour in the manner predicted by Sect.~\ref{sec:results-error_analysis}. Its substantially higher atmospheric metallicity increases the contribution of heavy molecules to the bulk atmospheric mass, raising the fractional uncertainty on $\mu$ to approximately $20\%$, while the uncertainty on C/O remains comparable to the other retrievals at approximately $55\%$.

\section{Discussion}\label{sec:discussion}

\subsection{Molecular redistribution under disequilibrium chemistry}
Water is the dominant radical source in warm H$_2$-rich atmospheres because it is both abundant and readily photodissociated at ultraviolet wavelengths. The resulting H and OH radicals initiate the oxidation network that controls much of the observable carbon chemistry. OH efficiently oxidises CO to \ce{CO2}, while H atoms drive methane oxidation through successive hydrogen-abstraction reactions. Although \ce{CH4} also photodissociates in the upper atmosphere, radical-driven chemistry dominates throughout most of the transmission region. In contrast, CO remained comparatively resistant to direct photolysis
because dissociation of its strong $\mathrm{C}\equiv\mathrm{O}$ triple bond
requires substantially higher photon energies than cleavage of the
$\mathrm{O-H}$ and $\mathrm{C-H}$ bonds in \ce{H2O} and \ce{CH4},
respectively: approximately $11.5\,\mathrm{eV}$ for
$\mathrm{C}\equiv\mathrm{O}$, compared with $5.9\,\mathrm{eV}$ for
$\mathrm{O-H}$ and $6.7\,\mathrm{eV}$ for $\mathrm{C-H}$.

Introducing nitrogen chemistry expands the set of observable carbon reservoirs rather than fundamentally changing the carbon chemistry. HCN provides an additional sink for carbon under reducing conditions, while N$_2$ remains chemically inert because of its strong triple bond. Sulfur follows a similar pattern, with most sulfur remaining locked in a small number of stable reservoirs (\ce{H2S} in the deep atmosphere; SH, \ce{S2}, \ce{CS2} and ultimately atomic S at lower pressures under photochemistry), and contributing little to either the observable oxygen chemistry or the bulk atmospheric mass. \changed{Therefore, these additional pathways modify} molecular partitioning without fundamentally altering the dominant carbon, oxygen, or hydrogen reservoirs.

\subsection{Mean molecular weight and the radius valley}
Our results are particularly relevant to recent proposals that atmospheric $\mu$ may vary systematically across the sub-Neptune radius valley. \citet{hengowentian2025} predicted such a gradient from geochemical outgassing models, with the trend controlled primarily by interior oxygen fugacity rather than atmospheric carbon enrichment. In this context, the relative robustness of $\mu$ found here is encouraging. If $\mu$ is less sensitive than elemental ratios to molecular abundance uncertainty and transmission-region photochemical redistribution, then population-level trends in $\mu$ may provide a more stable link between observable atmospheres and interior conditions. However, our models do not include interior outgassing or atmospheric escape, so they should be viewed as testing the atmospheric robustness of $\mu$, not the origin of any radius-valley trend. 

The analytic calculation assigns the same fractional uncertainty to every molecular abundance. Molecular constraints in atmospheric retrievals are not uniform\changed{:} feature strength, wavelength coverage, overlapping opacity, clouds, and model assumptions can cause one gas to be constrained more precisely than another. \changed{Therefore, the common fractional uncertainty provides} a controlled baseline rather than a realistic retrieval model. It isolates how the mathematical definitions of O/H, C/O, and mean molecular weight propagate a fixed level of molecular abundance uncertainty. Gas-dependent uncertainties could change the numerical contrast between these metrics, particularly when the least-constrained gas dominates an elemental inventory. The published case studies partially relax the common-uncertainty assumption by propagating the gas-dependent abundance uncertainties reported by each retrieval. These cases recover the same qualitative hierarchy between C/O and mean molecular weight. However, propagation of marginal abundance intervals does not capture covariance between retrieved gases.

\subsection{Uncertainty limits on C/O as a formation tracer}
The large propagated uncertainties in C/O introduce an observational
caveat to its use as a formation tracer. Disk-chemistry models
predicted that the C/O ratios of gas and solids vary across volatile
snowlines \citep{Oberg11}, while observations showed that disk C/O can
also vary spatially within individual planet-forming systems
\citep[e.g.][]{Bosman21,Keyte23}. Planet-formation calculations further
showed that atmospheric C/O depends not only on formation location but
also on the relative amounts of gas, pebbles, and planetesimals
accreted by the planet. Consequently, planets formed at different
orbital radii can acquire similar atmospheric C/O values, preventing
a unique formation location from being inferred from C/O alone
\citep{Cridland19,Bergin24}. Our results revealed a complementary
limitation: calculating C/O from uncertain molecular abundances can
amplify their uncertainties, even when C/O remains chemically
meaningful. Therefore, formation interpretations based on retrieved
C/O require chemical and disk-evolution modelling together with
careful propagation of the molecular abundance uncertainties.

\section{Conclusion}\label{sec:conclusion}

We developed an analytic framework to quantify how uncertainties in retrieved molecular abundances propagate into the derived atmospheric metrics C/O, O/H, and mean molecular weight, $\mu$. We evaluated these uncertainties across a sampled $T$--$P$--C/O grid, tested the predictions against photochemical forward models spanning a range of vertical mixing efficiencies, and applied the same approach to representative published atmospheric retrievals of hot Jupiters and sub-Neptunes.

The analysis shows that elemental abundance ratios are substantially more sensitive to molecular abundance errors than mean molecular weight. Photochemical forward models show that this relative ordering is preserved under disequilibrium chemistry. Although photochemistry and vertical mixing substantially redistribute molecular abundances, they produce only minor changes in the derived atmospheric metrics. The same qualitative behaviour is recovered in published atmospheric retrievals.

Our results indicate that the principal limitation on retrieved elemental abundance ratios arises from amplified molecular abundance uncertainties rather than from photochemical redistribution of molecular reservoirs. The method presented here provides a computationally inexpensive way to estimate these uncertainties and establishes a baseline for future investigations incorporating additional atmospheric processes, including condensation, cloud formation, and higher-metallicity atmospheres.

\begin{acknowledgements}
The authors thank the anonymous referee for constructive comments that
improved the clarity of the manuscript. B.B. acknowledges financial support from the European Research Council (ERC)
Geoastronomy Synergy Grant (grant No. 101166936; PIs: Gaillard, Heng, and
Mojzsis). The authors acknowledge the use of generative AI tools to assist with
language editing and manuscript refinement. All scientific content, analysis,
interpretation, and conclusions are the responsibility of the authors.
\end{acknowledgements}

%%%%%%%%%%%%%%%%%%%% REFERENCES %%%%%%%%%%%%%%%%%%

\bibliographystyle{aa}
\bibliography{mmw}

\end{document}